# A Computational Method to Simulate Electrostatic Actuation in Polycatenated Architected Materials

Mateus Maciel Vivaldi[1] and Alexandre F. Fonseca[1]*

[1] *Universidade Estadual de Campinas (UNICAMP), Instituto de Física Gleb Wataghin, Departamento de Física Aplicada, Campinas, São Paulo, 13083-859, Brazil.*

*Corresponding author: afonseca@ifi.unicamp.br

*ABSTRACT*

*A novel class of polycatenated architected materials (PAMs) has recently emerged, exhibiting distinctive mechanical properties such as stress-strain hysteresis and a geometry-driven transition between solid-like and fluid-like behavior. Although finite element models exist for conventional PAMs, none currently account for electrostatic effects. Here, we propose and qualitatively validate a computational framework to simulate the structure, dynamics, and mechanical response of electrostatically charged PAMs. Our approach successfully reproduces electrostatic actuation in close agreement with experiments, and reveals that electrostatic charges enable a reversible fluid-to-solid transition while also providing a means to estimate electrostatic stiffness. Notably, stiffness enhancements of up to sevenfold are achieved upon charging. This work expands the design space of PAMs and offers a predictive tool for actively controllable mechanical metamaterials.*

## INTRODUCTION

Mechanical metamaterials derive their unconventional properties – negative compressibility, zero or negative Poisson's ratios, negative thermal expansion, fluid-like responses, among others – from geometric design rather than composition [1–17]. Polycatenated architected materials (PAMs) constitute an emerging subclass, comprising mechanically interlocked rings whose topology is programmed via crystalline lattice placement of ring centers [18]. This design imparts tunable hysteresis and solid-fluid transitions, yet electrostatic effects remain unexplored within actual suitable simulation tools [19-21].

Here, we introduce a LAMMPS-based computational framework to model and qualitatively predict the structure, dynamics, and mechanical behavior of charged PAMs [22]. Each ring is represented as a rigid assembly of discrete point particles, enabling explicit, directionally resolved electrostatic interactions along the ring surface, a capability absent in prior granular-package approaches [23-25]. The model captures charge-induced stiffening and fluid-to-solid transitions, as demonstrated for two representative topologies.

The next sections detail the method, present the results, and summarize the main conclusions.

## THEORY AND SIMULATION DETAILS

LAMMPS is an open-source particle-based simulation package widely employed in atomistic, coarse-grained, and continuum-scale modeling [22]. Although computationally more demanding than mesh-based finite-element methods for certain mesoscopic systems, particle-based representations offer distinct advantages for problems involving nonlocal interactions, nonlinearities, and polydispersity – key features in the study of charged PAMs. The package numerically integrates Newton's equations of motion using force-field-derived potentials and includes tools for thermal and pressure control, enabling multiscale investigation of structural, mechanical, and dynamical behavior provided that the interactions capture the relevant physics.

Two PAM topologies, J4 and T6, as designated by Zhou *et al*. [18] are examined in this study. The numerical suffixes indicate the number of concatenated rings per ring, while the letters J and T refer to the NbO [26] and polybenzene [27] crystalline topologies, respectively. Figure **1** depicts the point-based ring representation, including dimensions and geometric parameters, along with unit cells and representative three-dimensional models of both architectures.

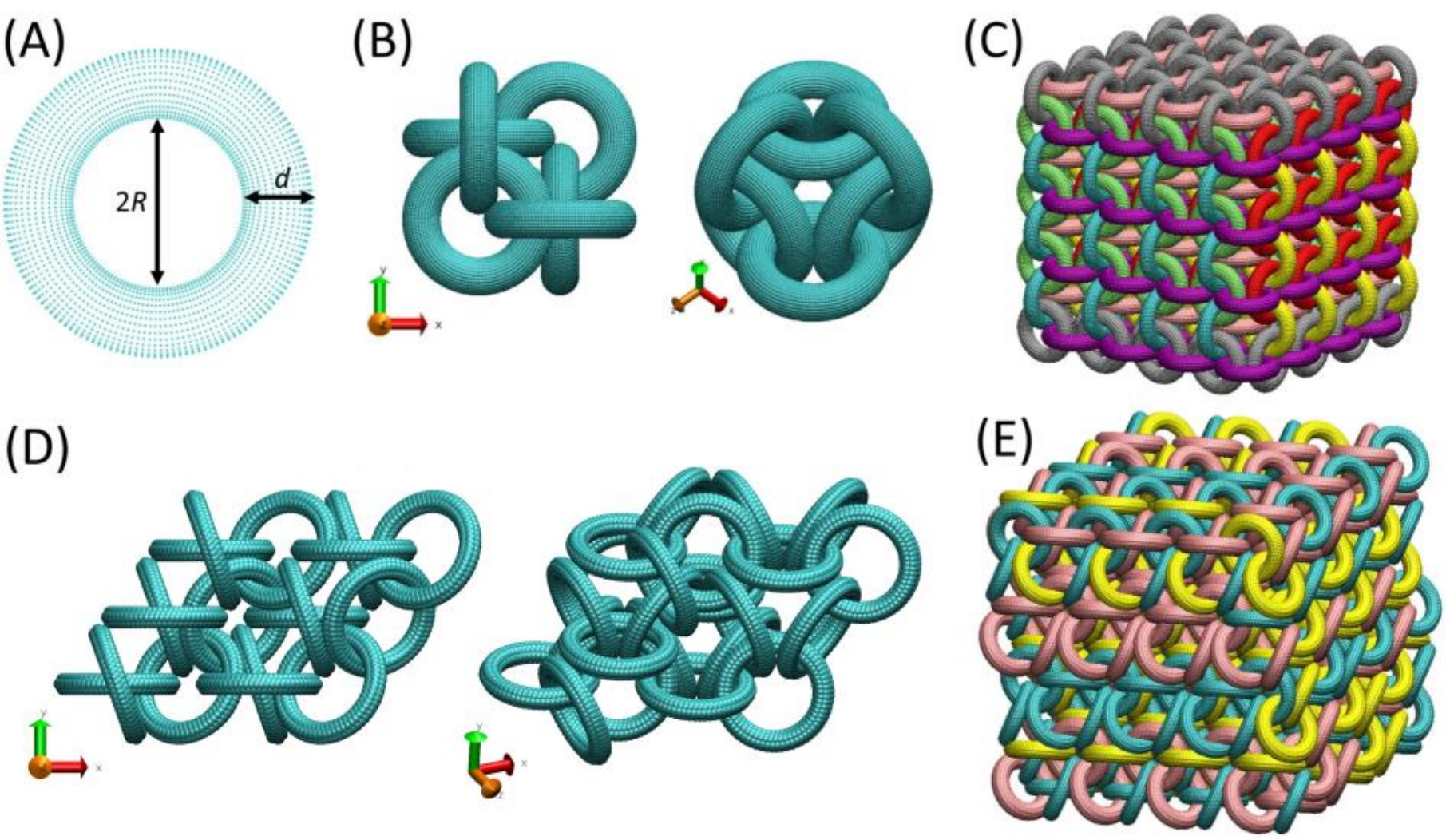

**Figure 1.** Circular ring discretization and PAM structures considered in this study. (A) Schematic of a single ring, defined by inner radius $R$ and thickness $d$. (B, D) Two views of the unit cells of the J4 and T6 topologies, respectively. (C, E) Corresponding supercells – J4 with 384 rings (952 points each) and T6 with 543 rings (690 points each) – assembled by periodic replication along $x$,$y$,$z$. Individual rings are color-coded for clarity in (C) and (E).

In our implementation, Lennard-Jones (LJ) and Coulomb interactions act exclusively between points of distinct rings, with cutoff truncation. Gravity is applied as a constant vertical force on all points, regardless of charge state. A Langevin thermostat at 0 K provides damping in lieu of explicit ring-ring friction, yielding a simple, but computationally efficient, framework for electromechanical studies.

Each ring consists of uniformly distributed surface points to ensure isotropic inter-ring interactions. The *fix rigid* command in LAMMPS simulates groups of points as rigid bodies. Because LAMMPS limits group definitions to 32, we circumvent this restriction by assigning each ring a unique group identifier via the function `ceil(id/Np)`, where `id` is the point index (1 to the total number of points) and `Np` is the number of points per ring. This yields a vector whose first *Np* entries equal 1, the next *Np* equal 2, and so forth, allowing the *fix rigid* command to distinguish individual rings. This approach requires the input file to be ordered such that every consecutive *Np* points correspond to a single ring.

Mass and charge are assigned uniformly to all points within a ring. Although possible, anisotropic distributions of mass and charge are deferred to future work. Thus, here, each ring is fully characterized by its total mass and charge.

Inter-ring repulsion is modeled via the repulsive branch of the 12-6 LJ potential, as implemented in LAMMPS:

$$U(r) = \begin{cases} \varepsilon \left(\frac{\sigma}{r}\right)^{12} & if \quad r < \sigma\,, \\ 0 & if \quad r \geq \sigma\,, \end{cases} \tag{1}$$

where $\varepsilon$ and $\sigma$ denote the well depth and equilibrium distance of the full LJ potential, respectively. The repulsion strength is governed by $\varepsilon$. This simplified form precludes the incorporation of material-specific properties, limiting our method to capturing mechanical effects arising solely from geometric and topological parameters. Coulomb interactions are truncated at 20 cm.

Compressive stress-strain simulations are performed by imposing planar walls at the top and bottom boundaries, with ring-wall interactions described by the same repulsive LJ potential (Eq. 1) using distinct $\varepsilon$ and $\sigma$ values. The bottom wall is fixed, while the top wall moves downward at constant velocity. For shear tests, the uppermost and lowermost rings are held rigid, and the upper group is translated horizontally at a constant rate. Preliminary shear runs establish the maximum strain each architecture can withstand – defined as the threshold before numerical instabilities (e.g., floating-point exceptions) arise – and subsequent simulations are restarted and run to that limit. Both protocols are initiated only after the system has reached gravitational equilibrium, which remains active throughout.

## RESULTS AND DISCUSSION

The results are organized into three complementary sets. The first comprises compressive and shear stress-strain simulations of the uncharged PAM structures depicted in Figure **1**, which are directly compared with the experimental data reported by Zhou *et al*. [18]. In particular, this analysis seeks to ascertain whether a single loading-unloading cycle reproduces the hysteresis characteristic of the experimental response. The second set offers a qualitative comparison between the electrostatic actuation of copper-coated J4 geometries – specifically, a "T"-shaped and a ladder-shaped configuration, as shown in Supplementary Movie S9 of Zhou *et al*. [18] – and our corresponding simulations (presented as Videos 1 and 2 in the Supplementary Information). The third set investigates the influence of electric charge insertion on the mechanical stiffness of two additional J4 and T6 PAM structures, with the goal of exploring the electrostatic stiffening phenomenon [28-30] through compressive stress-strain simulations.

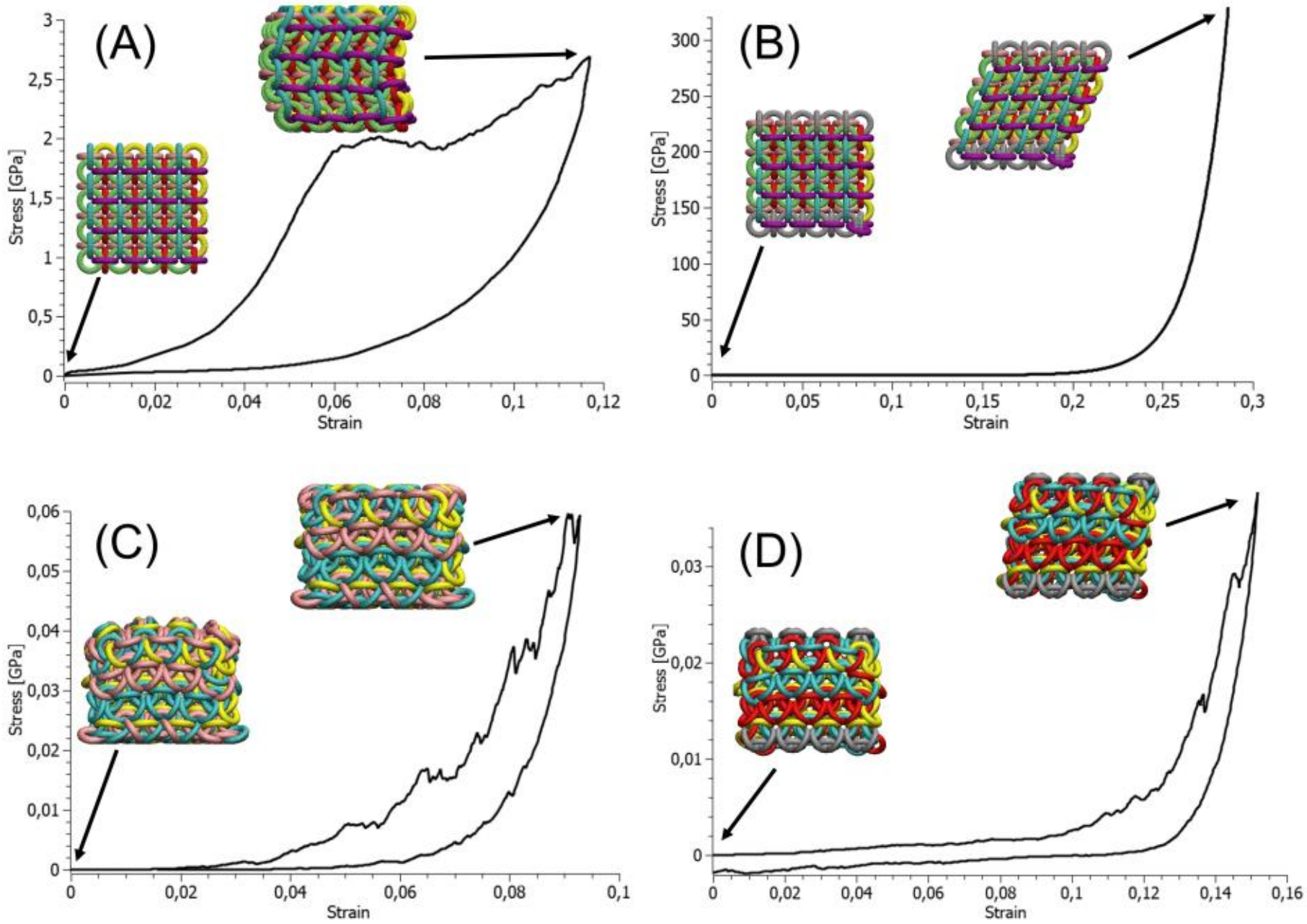

**Figure 2.** Compressive (A, C) and shear (B, D) stress-strain curves of the J4 (top) and T6 (bottom) PAM structures depicted in Figure **1**. Insets show representative snapshots of each structure at zero and maximum applied strain.

Figure **2** displays the compressive and shear stress-strain responses of the J4 and T6 PAM structures shown in Figure **1**. The J4 PAM is composed of 384 rings (with 952 points per ring), each with mass $m$ = 10 g, internal radius $R$ = 0.62 mm, and thickness $d$ = 0.37 mm. The T6 PAM consists of 543 rings (with 690 points per ring) of the same mass ($m$ = 10 g), but with larger characteristic dimensions ($R$ = 4.13 mm, $d$ = 1.79 mm). A comprehensive listing of all simulation parameters is provided in Table **1**.

**Table 1.** Parameters used to simulate the compressive and shear stress-strain relationships of J4 and T6 PAMs given in Figure **1**.

| Parameter | J4 | T6 |
|---|---|---|
| $\varepsilon$ (ring) | 1 mJ | 100 mJ |
| $\sigma$ (ring) | 0.09 mm | 0.35 mm |
| Timestep (compression) | 0.1 μs | 1 μs |
| Timestep (shear) | 0.1 μs | 0.1 μs |
| Compressive rate | 1 cm/s | 2 cm/s |
| Shear rate | 1 cm/s | 1 cm/s |
| Damping factor ($T_{\mathrm{DAMP}}$) | 10 μs | 10 μs |

| $\varepsilon$ (artificial wall) | 1 J | 1 J |
|---|---|---|
| $\sigma$ (artificial wall) | 0.89 mm | 0.50 mm |

The compressive stress-strain curves of the J4 and T6 PAMs exhibit hysteresis, consistent with experimental observations. While the absolute stress values carry no physical meaning owing to the limitations of the simulation method in capturing the intrinsic material properties of the ring constituents, the relative magnitudes of the peak stresses reflect the influence of topological differences on the overall mechanical response. Notably, the interaction parameter $\varepsilon$ prescribed in Table **1** is deliberately set two orders of magnitude higher for the T6 structure than for the J4 counterpart. Despite this, at comparable maximum compressive strains of approximately 10%, the peak stress recorded for the T6 structure is roughly two orders of magnitude lower than that of the J4 structure. This substantial reduction in strength is in full agreement with the established understanding that the T6 topology confers greater compliance under compressive loading [18].

The shear stress-strain curves present no hysteresis for the J4 structure and only a relatively small hysteresis for the T6 architecture, a trend that is qualitatively consistent with the experimental observations reported in [18]. In both cases, a regime of negligible mechanical response is observed up to characteristic strain thresholds – approximately 20% for J4 and 10% for T6 – delineating a fluid-like behavior at low deformations from a solid-like response at higher strains. Although our simulations do not reproduce the hysteresis loop for the J4 PAM, they successfully capture the correct ordering of the maximum sustainable strains, with J4 exceeding T6. Of particular note is the asymmetric behavior upon unloading: for the J4 structure, the stress paths during loading and unloading coincide at low strains, whereas for the T6 structure, a small but persistent deviation remains until the strain returns to zero. A similar unloading feature is visible in the shear response of a T6 sample presented in Figure 3F of Ref. [18].

Videos 1 and 2 in the Supplementary Information illustrate the electrostatic actuation simulations of two PAM structures configured in "T"-shaped and "ladder"-shaped geometries, analogous to those experimentally investigated by Zhou *et al.* [18]. As in the experiments, only the J4 topology is considered. Electric charges are introduced after the system has reached equilibrium under gravity alone. The electrostatic simulations are performed with $\sigma \cong 1.12$ mm, $\varepsilon = 0.01$ J, timestep = 1 μs and a Coulomb cutoff of 20 cm. The "T"-shaped structure is composed by 192 rings of 952 points, $R = 6.2$ mm, $d = 3.7$ mm, $m = 2$ g, and carried charge of 1 μC; actuation occurred over a timescale of 2 ms. The "ladder" configuration is composed of 540 rings of 100 points, $R = 2.68$ mm, $d = 0.80$ mm, $m = 1$ g, and a charge of 0.1 μC, with actuation completed in 80 ms.

Finally, to quantify the influence of inserted electric charges on the mechanical stiffness of the PAMs, we performed one cycle of compressive stress simulations on two cubic PAM architectures of distinct topologies, both with and without electrostatic charging. The J4 PAM comprises 162 rings of 100 points each ($R = 2.68$ mm, $d = 0.80$ mm, $m = 10$ g), whereas the T6 PAM consists of 198 rings of 95 points each ($R = 2.35$ mm, $d = 0.37$ mm, $m = 10$ g). All simulations were conducted with $\varepsilon = 0.01$ J, a timestep of 1 μs, and a Coulomb cutoff radius of 20 cm. Owing to the differences in ring thickness ($d$) between the two structures, the van der Waals parameter $\sigma$ were set to approximately 1.12 mm and 0.50 mm for the J4 and T6 PAMs, respectively. The resulting stress-strain curves are presented in Figure **3**.

We first draw attention to the equilibrium configurations attained under gravitational load. Panels (H) and (G) depict the uncharged structures at zero applied strain. In this state, the T6 architecture undergoes significant self-collapse, a consequence of its substantially larger inter-ring void fraction relative to the J4 topology. This morphological distinction under gravity is critical for interpreting the subsequent stress-strain behavior.

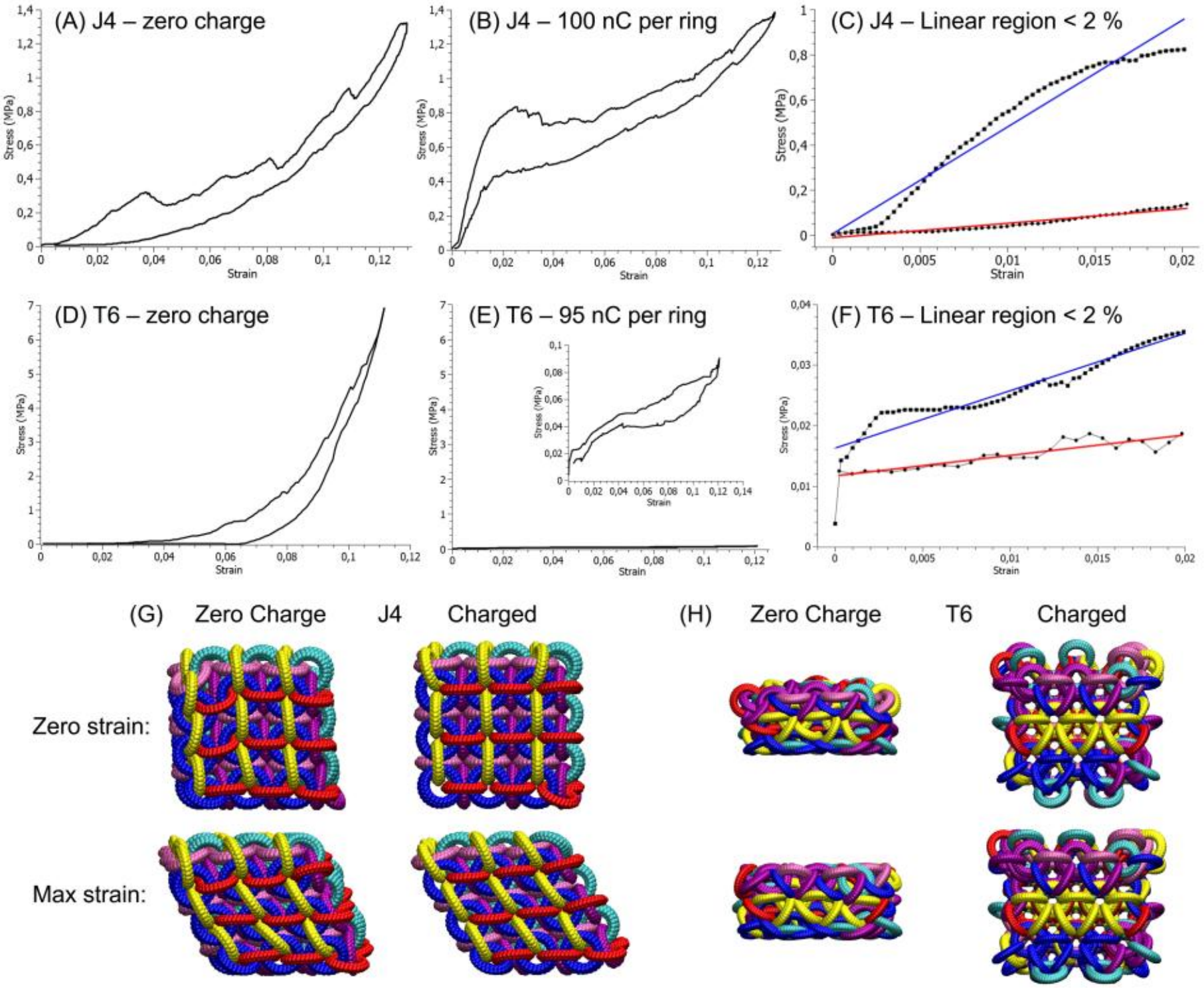

**Figure 3.** Mechanical response of J4 (162 rings, 100 points per ring) and T6 (198 rings, 95 points per ring) PAM structures under uniaxial compression. (A, D) Stress-strain curves for uncharged structures. (B, E) Corresponding curves with electrostatic charges inserted (1 nC per point). (C, F) Linear elastic regime (strain < 2%) comparing charged (squares) and uncharged (circles) conditions. Blue and red lines represent linear fittings. (G, H) Representative snapshots of J4 and T6 structures, respectively, at zero and maximum compressive strain, for both charge states. All rings have a mass of 10 g.

Figure **3** presents the compressive stress-strain response of both uncharged PAM structures over a full 10% strain cycle (panels A, D). Although the J4 structure is known to exhibit lower compliance, the T6 architecture sustains a peak stress approximately five times higher, an effect attributed to gravitational collapse of the T6 and concurrent shear deformation in the J4, consistent with experimental observations (Fig. **S8** in Ref. [18]).

Upon insertion of 1 nC per point (panels B, E), the J4 structure displays a marked stiffening at low strains, reflected in the linear elastic regime (strain < 2%, panel C). The extracted elastic modulus increases from 6.5 MPa to 44 MPa, a roughly sevenfold enhancement. While the absolute modulus values remain physically inconclusive due to the absence of explicit ring material properties, the clear trend underscores the tunability of electrostatic stiffness via charge injection.

The effect is even more pronounced for the T6 structure, which transitions from a fluid-like to a solid-like behavior solely upon charging (panel H). Despite a reduction in peak stress under the same maximum compressive strain, the charged T6 structure exhibits a distinct gain in stiffness, with its elastic modulus rising from 0.34 MPa (collapsed, uncharged) to 0.95 MPa (charged), an approximately threefold increase. This behavior indicates that charged T6 networks can accommodate substantially larger compressive strains than their neutral counterparts.

The observed mechanical divergence between the two topologies arises from geometric differences: the J4 structure has a $d/R$ ratio of $\cong 0.3$, versus $\cong 0.16$ for the T6. Architectures with smaller $d/R$ appear more susceptible to charge-induced stiffening, suggesting a geometry-dependent electrostatic response. A systematic exploration of the interplay between $d/R$, charging, and elastic modulus is deferred to future work.

## CONCLUSION

We have introduced a LAMMPS-based simulation framework for PAMs that, despite omitting explicit ring material properties, effectively captures the structural and mechanical consequences of electrostatic charging. The method accommodates arbitrary ring geometries, sizes, including nonuniform distributions of mass and charge, offering broad flexibility for future investigations.

Our simulations reveal that hysteresis in both compressive and shear stress-strain cycles arises even in frictionless PAMs, underscoring the dominant role of purely geometric and topological effects in their mechanical response. Furthermore, charge insertion enhances stiffness by up to sevenfold, demonstrating a viable route for tunable electromechanical design.

Rather than replacing conventional finite-element approaches, the proposed method serves as a complementary tool, particularly suited for probing charge-induced phenomena and structural reorganizations in complex PAM architectures.

## ACKNOWLEDGMENTS

This work used resources of the John David Rogers Computing Center (CCJDR) in the Gleb Wataghin Institute of Physics, University of Campinas. Computational resources were provided by the Coaraci Supercomputer for computer time (Fapesp grant #2019/17874-0) and the Center for Computing in Engineering and Sciences at Unicamp (Fapesp grant #2013/08293-7).

## FUNDING

AFF is a fellow the Brazilian Agency CNPq-Brazil (Grant number 302009/2025-6), and acknowledges São Paulo Research Foundation (FAPESP) (Grant numbers #2023/02651-0 and #2024/14403-4), and Fundação de Apoio ao Ensino, Pesquisa e

Extensão – FAEPEX/UNICAMP (Grant number # 3264//26). MMV acknowledges FAPESP Grant #2025/06328-5.

## AUTHOR CONTRIBUTION

A.F.F. conceived the idea and the research plan. A.F.F and M.M.V. performed the simulations and analyzed the results. A.F.F wrote the original draft and M.M.V. reviewed and edited the draft. A.F.F supervised the work and acquired funding. A.F.F. and M.M.V. read and approved the final manuscript.

## CONFLICT OF INTEREST STATEMENT

On behalf of all authors, the corresponding author states that there is no conflict of interest.

## DATA AVAILABILITY STATEMENT

Data available on reasonable request from the authors.